\documentclass{nature}
\usepackage[utf8]{inputenc}
\usepackage{graphicx}
\makeatletter
\let\saved@includegraphics\includegraphics
\AtBeginDocument{\let\includegraphics\saved@includegraphics}
\renewenvironment*{figure}{\@float{figure}}{\end@float}
\makeatother
\usepackage{dcolumn}
\usepackage{bm}
\usepackage{color,soul}
\definecolor{purple}{RGB}{127,0, 255}

\usepackage{caption}
\usepackage{ragged2e}
\usepackage{amssymb}
\usepackage{amsmath}
\usepackage{mathtools}
\usepackage{multirow}
\usepackage{lineno}
\renewcommand\st[1]{}

\title{3D nonlinear optical metamaterials from twisted 2D van der Waals interfaces} 
\author{Bumho Kim$^{1}$, Jicheng Jin$^{1}$, Zhi Wang$^{1}$, Li He$^{1}$, Thomas Christensen$^{2}$, Eugene J. Mele$^{1}$, Bo Zhen$^{1}$}

\begin{document}
\maketitle

\begin{affiliations}
 \item Department of Physics and Astronomy, University of Pennsylvania, Philadelphia, Pennsylvania 19104, USA
 \item Department of Physics, Massachusetts Institute of Technology, Cambridge, Massachusetts 02142, USA
\end{affiliations}

\section*{Abstract}

To enable new nonlinear responses, metamaterials are created by organizing structural units (meta-atoms) which are typically on the scale of about a hundred nanometers. However, truly altering atomic symmetry and enabling new nonlinear responses requires control at the atomic-scale, down to a few angstroms. Here we report the discovery of 3D nonlinear optical metamaterials realized by the precise control and twist of individual 2D van der Waals interfaces. Specifically, new nonlinear crystals are achieved by adding pseudo screw symmetries to a multiple of 4-layer WS$_2$ stacks (e.g. 4-layer, 8-layer, etc). The nonlinear susceptibility of the resulting 3D crystal is fundamentally different from natural WS$_2$. Furthermore, we show that the magnitude of the newly enabled nonlinearity is enhanced by controlling the number of interfaces and the excitation wavelength. Our findings suggest a new approach to redesigning intrinsic nonlinearity in artificial atomic configurations, scalable from a few nanometer-thick unit cells to bulk materials.

\section*{Introduction}

Material responses to light are highly constrained by symmetry. For example, two photons of identical frequency incident upon a material can excite second-harmonic (SH) polarization at doubled frequency with an amplitude proportional to the allowed second-order nonlinear susceptibility components of the material. Meanwhile, these nonlinear components are constrained by the point group symmetry of the materials\cite{boyd_nonlinear_nodate, sutherland_handbook_2003, bergfeld_second-harmonic_2003, wu_giant_2017, li_probing_2013}. Crystalline 3D materials have fixed intrinsic point group symmetries, which are often limited to naturally existing and stable phases of materials. Beyond natural materials, new nonlinear responses have been enabled in metamaterials through symmetry control\cite{kauranen_nonlinear_2012, lapine_colloquium_2014, li_nonlinear_2017, kadic_3d_2019}. For instance, breaking the inversion symmetry at the surfaces of natural materials\cite{shen_surface_1989, guyot-sionnest_local_1987, cazzanelli_second_2016} or large-scale ($\sim$100 nm) meta-atoms\cite{lee_giant_2014, butet_optical_2015, linnenbank_second_2016, li_nonlinear_2017} has enabled second-order nonlinear responses from inversion symmetric materials, where all second-order nonlinear susceptibility components are otherwise forbidden. Such new responses are mostly confined to the surfaces of materials, while the bulk mainly contributes to absorption losses\cite{li_nonlinear_2017, kadic_3d_2019, kauranen_nonlinear_2012, khurgin_how_2015}. As constituent meta-atoms inherently have low surface-to-volume ratios, current metamaterials typically have limited performances, such as overall frequency conversion efficiency, in practical applications\cite{khurgin_how_2015, kadic_3d_2019, bonacina_harmonic_2020, valev_characterization_2012}.

Different from conventional approaches in 3D materials, the advent of twisted van der Waals (vdW) materials enables control over their intrinsic point group symmetries\cite{wu_topological_2019, angeli__2021, zhang_electronic_2021, zhou_moire_2022}. The modified symmetry of twisted vdW stacks in turn significantly affects the electronic wavefunction at the twisted interfaces and enables previously forbidden new nonlinear responses\cite{yang_tunable_2020}. \textcolor{black}{While previous studies focus on controlling nonlinear responses at a single twisted interface\cite{hsu_second_2014, paradisanos_second_2022, van_der_zande_tailoring_2014, liu_evolution_2014, yang_tunable_2020} or a few twisted interfaces with a low interface-to-volume ratio\cite{yao_enhanced_nodate}, here we demonstrate a 3D crystal, completely made of twisted 2D interfaces.} In contrast to conventional metamaterials, our 3D crystal exhibit new \textcolor{black}{interfacial} nonlinear responses throughout its entire bulk\st{, promising highly efficient nonlinear conversion processes}. Overall, our study opens a new path to create and engineer 3D metamaterials by stacking and controlling individual 2D interfaces and promises new nonlinear functionalities for a broad range of applications including \textcolor{black}{freqeuncy conversion\cite{koshelev_subwavelength_2020, anthur_continuous_2020}}, bioimaging\cite{ray_size_2010, deka_nonlinear_2017}, ultrafast photonics\cite{hickstein_self-organized_2019, yu_high-yield_2021}, quantum computing\cite{kwiat_new_1995, zhang_electronic_2021}, and communication\cite{javerzac-galy_-chip_2016, fan_superconducting_2018}.

\section*{Theory for nonlinear twisted 3D crystals}
In this section, we introduce a new approach to building nonlinear 3D crystals from twisted 2D interfaces, as illustrated in Fig.~1. We start by discussing nonlinear responses of monolayer and twisted bilayer WS$_2$ based on their symmetries. A WS$_2$ monolayer belongs to the $D_{3h}$ point group, which supports second-order sheet susceptibility elements\cite{li_probing_2013, boyd_nonlinear_nodate, sutherland_handbook_2003} of $\chi^{(2)}_{yyy}=-\chi^{(2)}_{yxx}$ and enables an SH polarization along the armchair direction (denoted by the red arrow). Upon stacking two monolayers at a generic twist angle ($\phi \ne 0^{\circ}$ and $180^{\circ}$), the point group is reduced from $D_{3h}$ (monolayer) to $D_{3}$ (twisted bilayer) due to the breaking of mirror symmetries\cite{wu_topological_2019, angeli__2021, zhang_electronic_2021, zhou_moire_2022}. The twisted bilayers have both in-plane ($\chi^{(2)}_{xxx}=-\chi^{(2)}_{xyy}$) and interfacial ($\chi^{(2)}_{xyz}$ = - $\chi^{(2)}_{yxz}$) second-order susceptibility elements. While the former response (black arrow) is the sum of amplitudes from individual monolayer responses (red arrows), the latter response (blue arrow) is a new chiral response arising from broken mirror symmetries. Importantly, this interfacial nonlinear response does not come from either layer separately but is a cooperative effect originating from the coupling between the electronic wavefunctions from the two layers\cite{yang_tunable_2020}. 

Further analysis of this interfacial nonlinear susceptibility $\chi^{(2)}_{xyz}$ reveals two interesting properties. First, when the sample rotates in-plane, the interfacial SH polarization always points to a fixed direction, whereas the in-plane (and monolayer) SH polarization rotates with the sample (Supplementary Fig.~1). Furthermore, $\chi^{(2)}_{xyz}$ strongly depends on the relative twist angle of the top layer with respect to the bottom ($\phi$). The sign of $\chi^{(2)}_{xyz}$ flips when the twist angle is reversed: $\chi^{(2)}_{xyz}(\phi)$ = $-\chi^{(2)}_{xyz}(-\phi)$. See Supplementary Fig.~3 for more details and Fig.~2e for our experimental demonstration.

Twisted trilayer samples can be similarly analyzed as two twisted interfaces: one between the bottom layer and the middle, and the other between the middle and the top.  For example, a trilayer stack in the $\langle$0$^{\circ}$, 30$^{\circ}$, 60$^{\circ}\rangle$ configuration has two twisted interfaces, both at +30 degrees, creating identical interfacial SH polarizations that constructively add up. On the other hand, a trilayer stack in the $\langle$0$^{\circ}$, 30$^{\circ}$, 0$^{\circ}\rangle$ configuration has two oppositely twisted interfaces, one at +30 and one at -30 degrees. The opposite interfacial SH polarizations cancel out any chiral responses, which is consistent with the preserved up-down mirror symmetry of the stack.

Notably, screw symmetry emerges in a twisted four-layer stack in the $\langle$0$^{\circ}$, 30$^{\circ}$, 60$^{\circ}$, 90$^{\circ}\rangle$ configuration. Specifically, all interfacial SH polarizations (blue arrows) add up, whereas in-plane SH polarizations from individual layers (red arrows) exactly cancel between the first (second) layer and the third (fourth) layer as they point in opposite directions. Altogether, the four-layer stack only allows chiral interfacial susceptibility ($\chi^{(2)}_{xyz}$) that is different from the $D_3$ point group constraints in twisted bilayers and trilayers (Supplementary Table~1). This emerging constraint can be understood by a pseudo-$4_3$ screw symmetry that forbids in-plane susceptibility, $\chi^{(2)}_{xxx}$ (Supplementary Section I). Such an artificially added screw symmetry opens new paths to engineer 3D symmetries and nonlinearities of vdW stacks (see Supplementary Section II).

The unconventional nonlinearities that we discovered in twisted four-layer stacks are preserved in the 3D bulk crystal limit, which is well understood by the new point group. Specifically, a twisted bulk with a twist angle of $\phi=30^{\circ}$ between any two adjacent layers belongs to the $D_{12}$ point group. Applying Neumann's principle to the less understood $D_{12}$ point group, our theoretical analysis shows that only chiral susceptibility components, $\chi^{(2)}_{xyz}$, are allowed, while all in-plane susceptibility components (e.g., $\chi^{(2)}_{yyy}$ and $\chi^{(2)}_{yxx}$) are forbidden. This feature is consistent with a single four-layer twisted stack as we analyzed before. The consistency can be intuitively understood as the bulk crystal is essentially a vertical stacking of the 4-layer twisted stack. In other words, the 4-layer twisted stack is the new unit cell of the twisted bulk crystal, so they support the same nonlinear susceptibility components. 

\section*{Experimental demonstration}
We perform second harmonic generation (SHG) measurements to verify predicted nonlinearities of twisted WS$_2$ in the various stacking configurations. Twisted WS$_2$ stacks, up to eight-layer thick, are prepared on fused silica substrates using the ``tear and stack" method\cite{kim_van_2016, cao_correlated_2018} (see Methods). In contrast to SHG under normal excitation only arising from in-plane responses, we measure SHG under the oblique incident angles ($\beta$) set to be either +50$^{\circ}$ and -50$^{\circ}$ to observe both in-plane and interfacial nonlinear responses (Fig.~2a and Methods for details). A monolayer shows identical SHG signals between $\beta=50^{\circ}$ (red circles, Fig.~2b) and $-50^{\circ}$ (blue circles). This result indicates that the in-plane SH responses from the monolayer are independent of the opposite incident angle, which is also consistent with theoretical modeling based on the $D_{3h}$ point group (solid line, Fig.~2b).

In contrast, twisted bilayers, belonging to the $D_3$ point group, create SH polarizations in two ways: through the in-plane layer responses of $\chi^{(2)}_{xxx}$ and the interfacial responses of $\chi^{(2)}_{xyz}$ producing a net SHG intensity.
\begin{equation}
I_{\rm SHG} \propto (-\chi^{(2)}_{xxx}\cos^2\beta (\cos2\alpha \cos3\theta + \cos\beta \sin2\alpha \sin3\theta) + \chi^{(2)}_{xyz}\sin2\beta \cos2\alpha)^2.    
\label{eq.1}
\end{equation}
Here $\theta$ represents the in-plane sample orientation with respect to the lab frame.
The SHG intensity is proportional to the square of SH polarizations from individual layers (the $\chi^{(2)}_{xxx}$ term in Eq.~1) and the interface (the $\chi^{(2)}_{xyz}$ term). Unlike monolayer, the observed SHG signal is generally stronger for $\beta=50^{\circ}$ than $\beta=-50^{\circ}$ for the $\langle0^{\circ}, -30^{\circ}\rangle$ bilayer stack (Fig.~2c); conversely, the opposite trend is observed for the $\langle0^{\circ}, 30^{\circ}\rangle$ bilayer stack (Fig.~2d). Specifically, under the two oblique incidences ($\beta=\pm50^{\circ}$), the in-plane SH polarizations ($\chi^{(2)}_{xxx}$) are identical, whereas the interfacial SH polarizations ($\chi^{(2)}_{xyx}$) are exactly opposite \textit{cf.} Eq.~1.
The clear difference in SHG signals under the two opposite incident angles evinces the interfacial response of the twisted bilayers.

Next, we examine the interfacial responses of bilayer samples with opposite twist angles, $\phi = \pm 30^{\circ}$ in more detail. Following Eq.~1, the difference between SHG signals, $\Delta I_{\rm SHG} = I(\beta = +50^{\circ}) - I(\beta = -50^{\circ})$, is proportional to the interfacial susceptibility $\chi^{(2)}_{xyz}$ (Supplementary Section IV). The $\langle0^{\circ}, 30^{\circ}\rangle$ (green circles, Fig.~2e) and $\langle0^{\circ}, -30^{\circ}\rangle$ (orange circles) bilayer stacks have $\Delta I_{\rm SHG}$ of approximately equal magnitude but opposite signs. Using our theoretical model (green and orange solid lines), we extract $\chi^{(2)}_{xyz, \rm norm} = 0.021$ for the $\langle0^{\circ}, 30^{\circ}\rangle$ stack and $\chi^{(2)}_{xyz, {\rm norm}} = -0.024$ for the $\langle0^{\circ}, -30^{\circ}\rangle$ stack. Both values are normalized to the monolayer susceptibility (see Supplementary Section V). This result is consistent with our earlier prediction, $\chi^{(2)}_{xyz}(\phi)=-\chi^{(2)}_{xyz}(-\phi)$, namely the interfacial SH polarization reverses its direction when the twist angle is reversed as shown in Fig.~2f.

We next examine the scalability of interfacial responses by measuring a trilayer stack in the $\langle$0$^{\circ}$, 30$^{\circ}$, 60$^{\circ}\rangle$ configuration. Fig.~3a shows a schematic of interfacial SH polarizations (blue arrows) pointing to the same direction and adding up. The observed SHG signals from the trilayer show a significant difference between $\beta=50^{\circ}$ and $-50^{\circ}$, indicating finite interfacial responses (Fig.~3c,e). The observed $\Delta I_{\rm SHG}$ (circles, Fig.~3e) is consistent with our theoretical model (solid line), which yields $\chi^{(2)}_{xyz, {\rm norm}} \approx 0.032$, exceeding the corresponding response in the $\langle0^{\circ}, 30^{\circ}\rangle$ bilayer stack ($\chi^{(2)}_{xyz, \rm norm} = 0.021$). This experimental observation agrees well with our theory prediction: the interfacial nonlinear responses coherently add up when interfacial twist angles are the same. \textcolor{black}{This stacking sequence is different from widely studied 3R type stacks ($\phi=0^{\circ}$) where the in-plane layer nonlinear responses coherently add up\cite{yao_enhanced_nodate, liu_disassembling_2020}.} To further validate the origin of the differences in SHG, we examine a trilayer in the $\langle$0$^{\circ}$, 30$^{\circ}$, 0$^{\circ}\rangle$ configuration with two oppositely twisted interfaces (Fig.~3b). Fig.~3d,e show the SHG signals with almost no difference at the incident angles between $\beta=50^{\circ}$ and $-50^{\circ}$, implying a negligible net interfacial response: $\chi^{(2)}_{xyz} \approx 0$, which results from the cancellation between the two opposite interfacial SH polarizations as we predicted. The contrasting results between two trilayers with the same constituent layers but different interfaces confirm the observed $\Delta I_{\rm SHG}$ originates from vdW interfaces.

We experimentally demonstrate the unconventional susceptibility ($\chi^{(2)}_{xxx}=0$ and $\chi^{(2)}_{xyz}\ne0$) of a twisted four-layer unit cell structure. We first measure the in-plane response, $\chi^{(2)}_{xxx}$, of the four-layer stack under normal excitation ($\beta=0^{\circ}$), where we observe a substantially reduced response (purple, Fig.~4b) relative to the monolayer (grey). This strong suppression of $\chi^{(2)}_{xxx}$ is consistent with the presence of the pseudo $4_3$ screw symmetry (Fig.~4a). Fig.~4c shows stronger SHG signals in oblique excitation measurements ($\beta=\pm 50^{\circ}$) compared to the normal incident measurement ($\beta= 0^{\circ}$). This is direct evidence of stronger interfacial susceptibility than in-plane susceptibility in the twisted four-layer stack, which agrees with our previous theoretical prediction. \textcolor{black}{Moreover, the nearly identical SHG signals at the two opposite incident angles are consistent with the susceptibility of the four-layer stack: $\chi^{(2)}_{xxx}=0$ and $\chi^{(2)}_{xyz}\ne0$ \textit{cf.} Eq.~1.} Notably, the intrinsic nonlinearities of the four-layer stack are the same as those of some 3D materials (e.g. TeO$_2$\cite{okada_measurement_1977} and La$_4$InSbS$_9$\cite{zhao_strong_2012}) but fundamentally different from those of previously studied natural and twisted 2D materials (see Supplementary Table~2). The nonlinear susceptibility in the four-layer stack is accurately engineered by controlling the electronic wavefunction symmetries through twists, exemplifying a new type of nonlinear optical metamaterials.

To scale up a four-layer unit cell into a 3D crystal, we experimentally verify the nonlinearities of a vertical stack of two 4-layer unit cells (Fig.~4d). Our experimental results show a strong suppression of $\chi^{(2)}_{xxx}$ (Fig.~4e) and an even further enhancement of $\chi^{(2)}_{xyz}$ (Fig.~4f) in the twisted 8-layer stack, consisting of two 4-layer unit cells. Although strongly suppressed, $\chi^{(2)}_{xxx}$ is increased relative to the four-layer stack, which might be caused by the reabsorption of SHG from each layer as well as imperfect twist-angle alignment (see Supplementary Section VI and Supplementary Table~3). Meanwhile, $\chi^{(2)}_{xyz}$ scales approximately linearly with sample thickness, which is consistent with our theoretical model incorporating interference and reabsorption effects (solid line). Our observation promises even stronger interfacial responses in thicker samples by simply stacking 4-layer unit cells together. Moreover, the interfacial nonlinearity can also be further enhanced by the excitonic resonance effect (Fig.~5). The in-plane susceptibility of a monolayer has two prominent peaks at A-exciton ($\sim$620 nm) and B-exciton wavelengths ($\sim$525 nm), known as the exciton enhanced SHG\cite{fan_mechanism_2020, seyler_electrical_2015}. The excitonic resonance effect is also observed in the interfacial susceptibility of four-layer WS$_2$, reaching $\chi^{(2)}_{xyz, {\rm norm}}\sim 0.14$, which is over 3 times higher than the off-resonance value.

\section*{Conclusion}
To conclude, we show a new method of enabling nonlinear optical metamaterials, scalable from a twisted four-layer to a twisted bulk. We engineer the symmetries of electronic wavefunctions in 3D structures by controlling individual atomic-thick layers, which substantially reduces the size of optical metamaterials down to a few nanometers. Our new 3D crystals show completely redesigned nonlinear susceptibility components, predominantly originating from 2D interfaces, and take us one step closer to designing intrinsic nonlinearities at will. The enabled nonlinearities can be further enhanced by increasing the sample thickness and taking advantage of the exciton resonances. Moreover, we show that an artificially added screw symmetry can substantially modify the point group of 3D vdW materials, for example into a quasi-crystal, which provides a practical platform to explore emerging nonlinearity. Our approach to controlling symmetries can be readily extended to other physical responses that are also sensitive to material symmetries including elasticity, thermal expansion, and piezoelectricity\cite{newnham_properties_2004, hu_symmetry_2000, duerloo_intrinsic_2012, hu_localized_2018}.

\clearpage
\section*{Methods}
\subsection{Sample preparation}
A gold tape was prepared following the detailed procedures in the previous study\cite{liu_disassembling_2020}. A 150 nm-thick gold film on SiO$_2$/Si substrates was coated with polyvinylpyrrolidone (PVP) and then picked up by a thermal release tape. A freshly cleaved surface of the gold film was attached to a freshly cleaved WS$_2$ bulk crystal (CVT crystal from HQ graphene). Au exfoliation of a large ($\sim\,1\times1\,\mathrm{cm}^2$) bulk WS\textsubscript{2} crystal produces various sizes of continuous single-crystal monolayers up to near the lateral size of the bulk crystal. A gold tape attached on top of a polydimethylsiloxane (PDMS) microlens was used to pick up a part of a single-crystal WS$_2$ monolayer and reposition it on top of the remaining monolayer after adjusting the twist angle using the ``tear and stack" method\cite{kim_van_2016, cao_correlated_2018}. This procedure was repeated to produce up to an eight-layer WS$_2$ stack. A $\langle$0$^{\circ}$, 30$^{\circ}$, 60$^{\circ}$, 90$^{\circ}\rangle$ four-layer WS\textsubscript{2} stack was prepared by picking up part of a bilayer WS$_2$ stack with internal twist angle of 30$^{\circ}$ and placing it on top of the remaining bilayer after 60$^{\circ}$ rotation. Similarly, we fabricate a 2-unit cells eight-layer stack by lifting and stacking a four-layer without rotation. See Supplementary Fig.~6 for optical microscope images of the four-layer and eight-layer stacks. Twisted angles of each stack are estimated by polarization-resolved SHG, showing the uncertainty of $\sim\pm 1^{\circ}$ from the target twist angles.

\subsection{Details of SHG measurement}
An optical parametric oscillator (OPO from Light Conversion) with a wavelength of 1030 nm, repetition frequency of 75 MHz, and 96 fs pulse duration was used to measure SHG responses for all the samples. SHG signals of twisted bilayer (Fig.~2), trilayer (Fig.~3), and four-layer stacks (Fig.~4b,c) were acquired using the setup in Fig.~2a. The sample was mounted on a rotational stage, allowing us to tilt the sample around the $x$-axis, and thus to control the incidence angle ($\beta$). Vertically polarized OPO light was focused onto the tilted sample through an apochromatic 10$\times$ objective lens with a numerical aperture (NA) of 0.26 (Mitutoyo). The transmitted SHG signals were collected by the same second objective lens, passed through a HWP and horizontal polarizer, and are finally directed to a thermoelectrically cooled 2-dimensional (2D) charge-coupled device (CCD) array (iKon-M 912, Andor Technology) equipped with a spectrometer (Kymera 328u, Andor Technology). 

For four-layer and eight-layer stacks, imperfect sample preparation can exacerbate distortions of the twist angle. To minimize spatial inhomogeneity, a second setup with a reflective objective lens was used to estimate the normalized $\chi^{(2)}_{xyz}$ of the four- and eight-layers (purple squares in Fig.~4f). Transmission SHG is performed using a reflective objective lens (LMM40X-UVV from Thorlabs) with a NA of 0.5, which \textcolor{black}{enables simultaenous injections of the two oblique incidences at $\beta=\pm22.5^{\circ}$ (Supplementary Fig.~5a). The oblique beams are tightly focused onto a spot at a diffraction-limited radius of $\sim$ 1.3 $\mu {\rm m}$.}\st{enabled oblique injections of the OPO light by $\beta=\pm22.5^{\circ}$ onto the sample. The focused beam has a diffraction-limited radius of $\sim$ 1.3 $\mu {\rm m}$.} The transmitted SHG signals were collected by an apochromatic 20$\times$ objective lens with a NA of 0.4 (Mitutoyo). The SHG signals created by p-polarized injection were selectively collected using a vertically aligned one-dimensional slit and read by the 2D CCDs (Supplementary Fig.~5b,d). \textcolor{black}{The difference in the SHG signals (Supplementary Fig.5c,e) is fitted with Eq.~(S12), yielding the interfacial susceptibility of the four-layer and eight-layer stacks (Fig. 4f) while minimizing the inhomogeneity effect.}

To estimate the excitonic resonance effect of second-order susceptibility elements, we performed SHG measurements using an optical parametric amplifier with a repetition frequency of 3 kHz and a pulse duration of $\sim$180 fs while tuning the excitation wavelength from 1030 nm to 1360 nm. The in-plane susceptibility ($\chi^{(2)}_{yyy}$) of a monolayer is estimated from SHG under normal excitation. The interfacial susceptibility ($\chi^{(2)}_{xyz}$) of a four-layer is estimated by measuring the difference in SHG ($\Delta I_{\rm SHG}/\chi^{(2)}_{xxx} \propto \chi^{(2)}_{xyz}$) using the setup with a reflective objective lens. Both susceptibility elements are normalized by $\chi^{(2)}_{yyy}$ of a monolayer at an excitation of 1030 nm (Fig.~5).

\subsection{Data availability}
The data within this paper are available from the corresponding author upon request. Source data are provided along with this paper.

\subsection{Acknowledgments}
This work was partly supported by the National Science Foundation through the University of Pennsylvania Materials Research Science and Engineering Center DMR-1720530, the U.S. Office of Naval Research (ONR) through grant N00014-20-1-2325 on Robust Photonic Materials with High-Order Topological Protection and grant N00014-21-1-2703, and the Sloan Foundation. Work by E.J.M is supported by the Department of Energy under grant DE-FG02-84ER45118.

\subsection{Author contributions} B.Z. and B.K. conceived the project. B.K. and Z.W. fabricated twisted stacks. B.K. performed the SHG measurements assisted by J. J., and L. H. B.K. and T.C. performed the symmetry analysis. B.Z., E.J.M, and B.K. discussed and interpreted the results. B.Z. and B.K. wrote the paper with input from all authors. All authors discussed the results.

\subsection{Competing interests} The authors declare no competing interest.
 
\subsection{Correspondence} Correspondence should be addressed to B.Z. (email: bozhen@sas.upenn.edu).

\clearpage

\begin{figure}%
\centering
\includegraphics[width=1\textwidth]{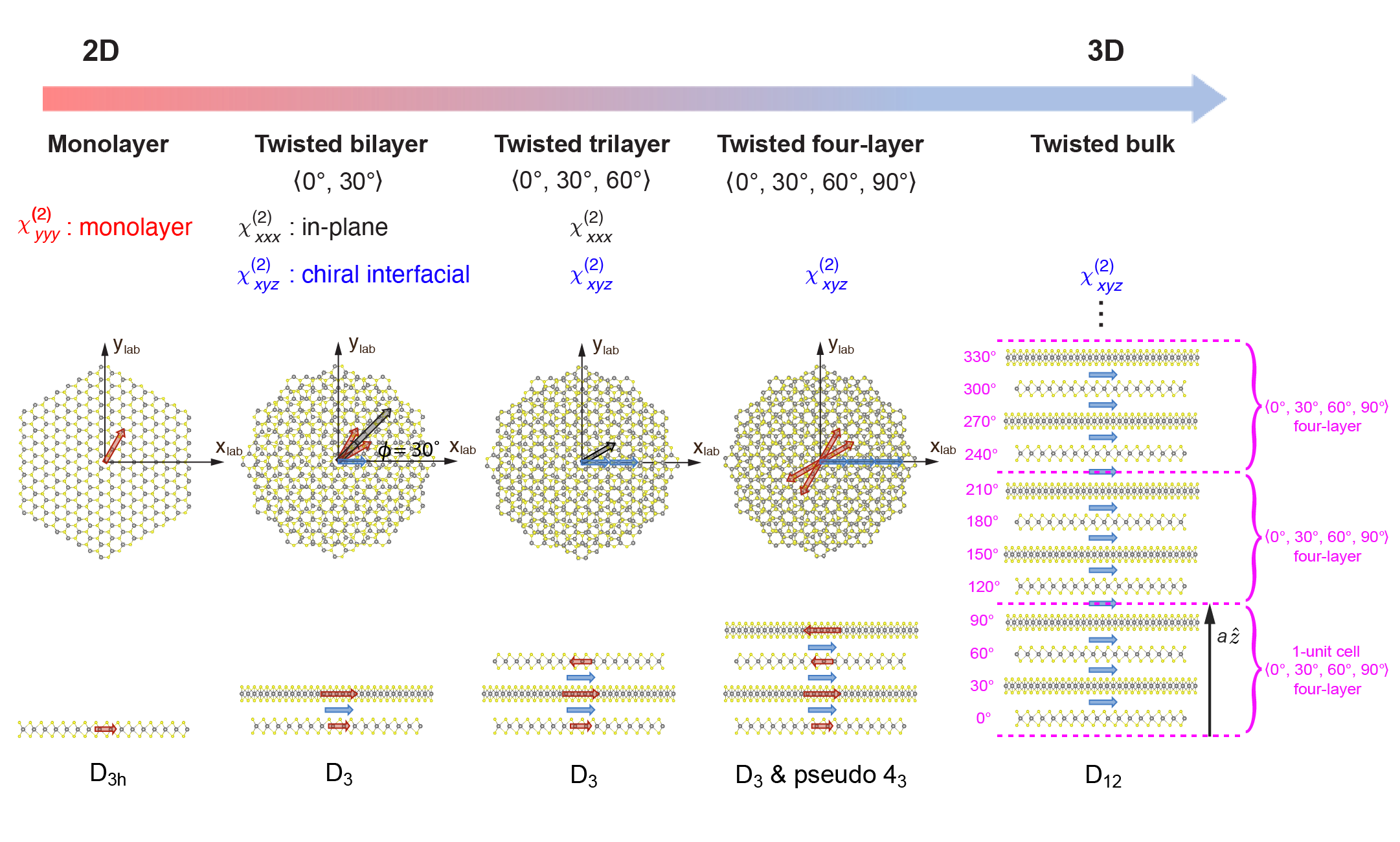}\\
\noindent\justifying\textbf{Fig.~1 $|$ Redesigned second-order susceptibility in a twisted 3D crystal.} Atomic structures and SH polarizations in different stacks, including a WS$_2$ monolayer ($D_{3h}$ symmetry), a $\langle 0^{\circ}, 30^{\circ} \rangle$ twisted bilayer stack ($D_3$ symmetry), a $\langle 0^{\circ}, 30^{\circ}, 60^{\circ} \rangle$ twisted trilayer stack ($D_3$ symmetry), a $\langle 0^{\circ}, 30^{\circ}, 60^{\circ}, 90^{\circ}\rangle$ four-layer stack ($D_3$ symmetry and pseudo $4_3$), and a twisted bulk with a fixed twist angle of 30$^{\circ}$ ($D_{12}$ symmetry) under $y_{\rm lab}$- and $z_{\rm lab}$-polarized excitation. The twisted bulk structure is equivalent to $\langle 0^{\circ}, 30^{\circ}, 60^{\circ}, 90^{\circ}\rangle$ four-layers stacked along the $\hat{\mathbf{z}}$ direction. Relevant allowed and forbidden second-order susceptibility elements ($\chi^{(2)}_{ijk}$) by symmetry are listed for each stack. Twisted four-layer and bulk only allow chiral interfacial nonlinear susceptibility ($\chi^{(2)}_{xyz}$) whose SH polarizations are denoted by blue arrows. SH polarizations along an armchair direction of each monolayer are denoted by red arrows whose sum is denoted by the black arrow. Grey and yellow circles represent tungsten and sulfur atoms, respectively.
\label{Fig1}
\end{figure}

\begin{figure}[h]%
\centering
\includegraphics[width=0.77\textwidth]{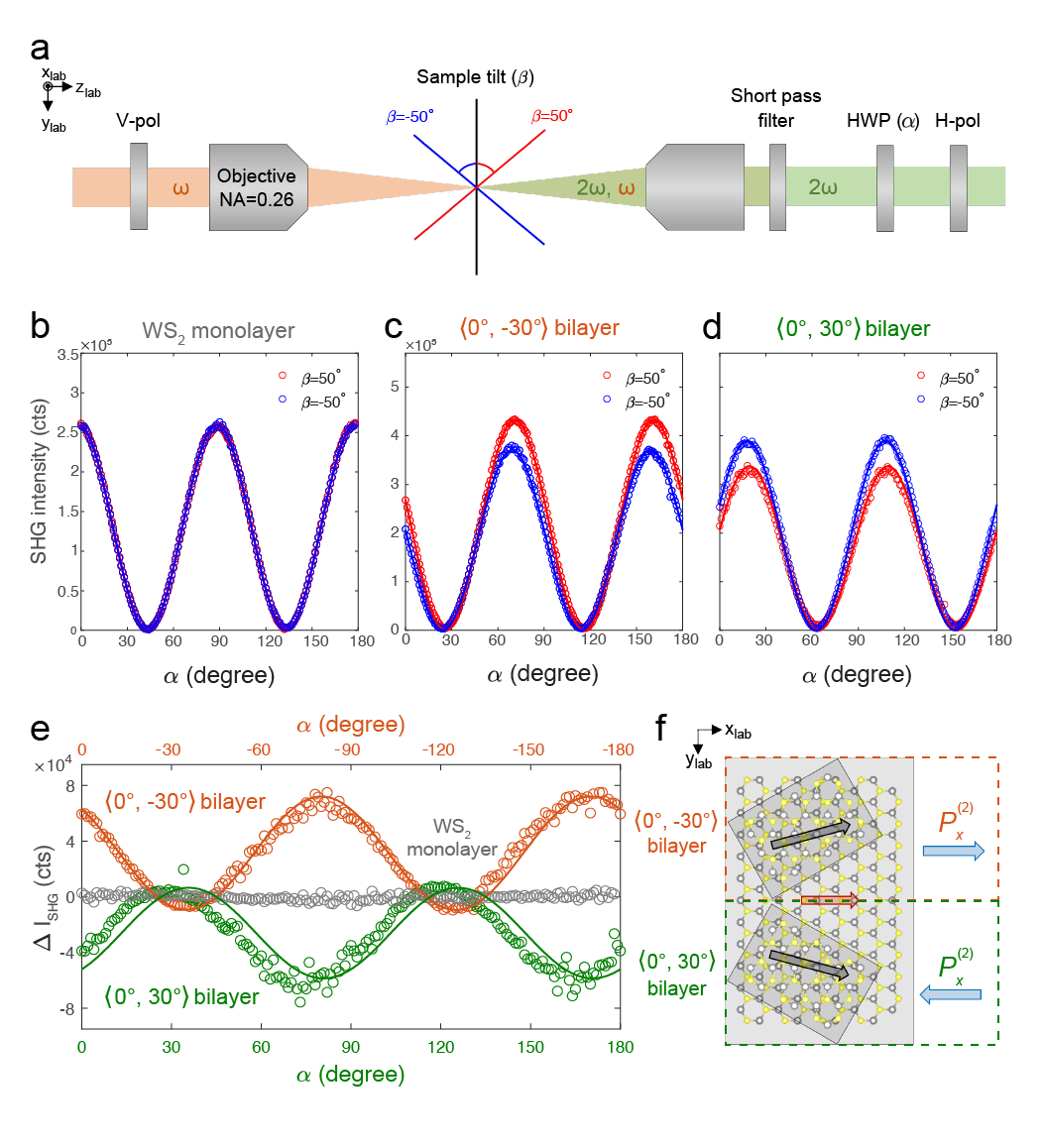}\\
\noindent\justifying\textbf{Fig.~2 $|$ Interfacial nonlinear susceptibility in twisted bilayer WS$_2$. a,} SHG measurement setup to characterize both individual layer and interfacial nonlinear responses. See Methods for more details. V-pol (H-pol) denotes vertical (horizontal) polarizer along the $y$($x$) direction. 
SHG signals are measured at the two incidence angles, $\beta=50^{\circ}$ in red and $\beta=-50^{\circ}$ in blue, on different devices, including \textbf{b}, monolayer, \textbf{c}, $\langle 0^{\circ}, 30^{\circ}\rangle$ and \textbf{d}, $\langle 0^{\circ}, -30^{\circ}\rangle$ bilayer stacks. 
Half-wave plate (HWP) orientation angle is denoted as $\alpha$.
\textbf{e}, Differences in the SHG signals ($\Delta I_{\rm SHG}$) between the incidence angles of $\beta=\pm50^{\circ}$ of a monolayer (grey), a $\langle 0^{\circ}, 30^{\circ} \rangle$ bilayer (green), and a $\langle 0^{\circ}, -30^{\circ}\rangle$ bilayer (orange circles). Solid lines in (\textbf{b--e}) are the best-fit curves from symmetry analysis (Eq.~1). \textbf{f}, Schematic of $\langle 0^{\circ}, \pm30^{\circ}\rangle$ bilayer stacks, mirror symmetric to each other, inducing oppositely aligned interfacial SH polarizations ($P_{x}^{(2)}$, blue arrows). 
\label{Fig2}
\end{figure}

\begin{figure}[h]%
\centering
\includegraphics[width=0.5\textwidth]{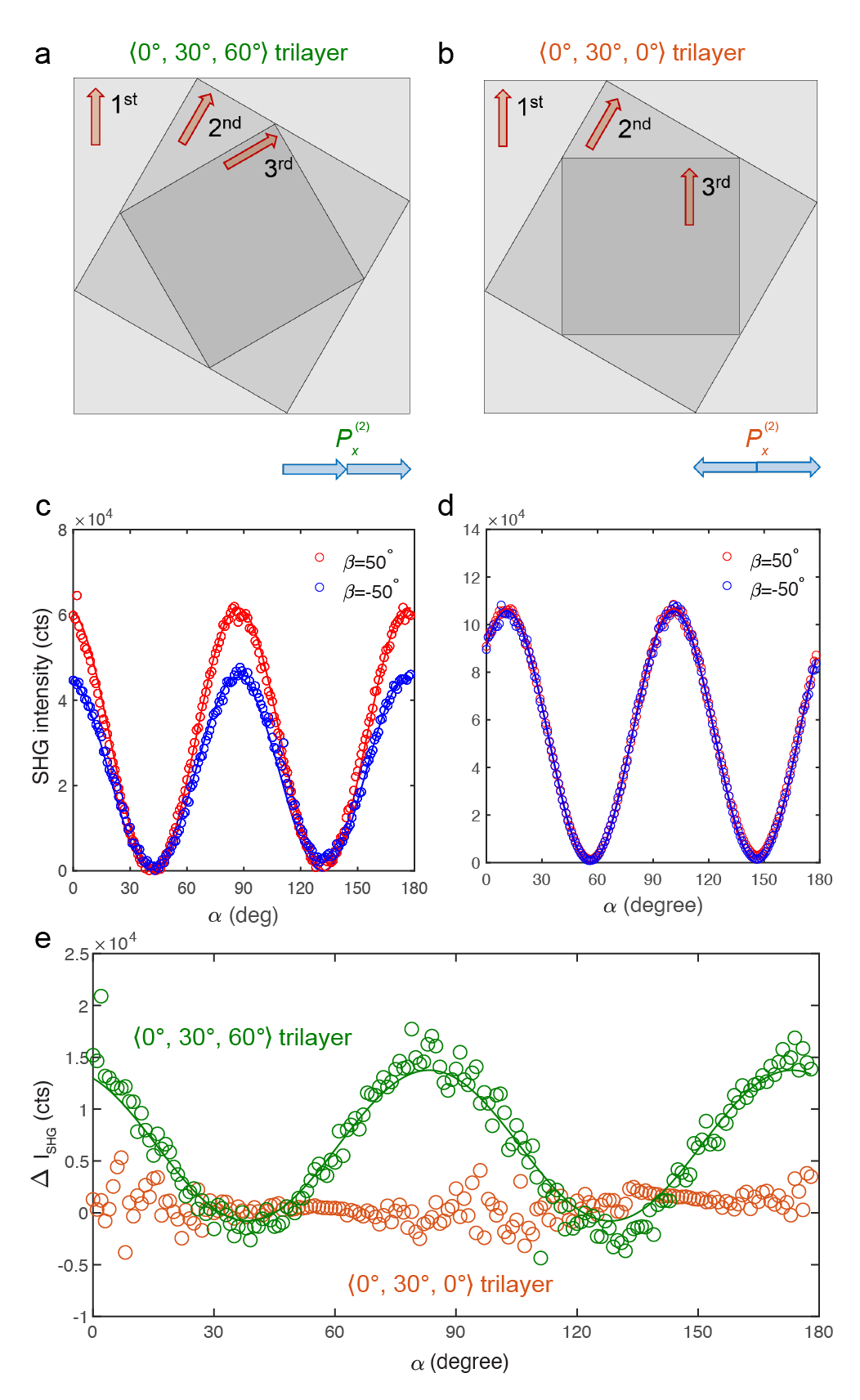}\\
\noindent\justifying\textbf{Fig.~3 $|$ Scalability of interfacial nonlinear susceptibility in twisted trilayer WS$_2$ stacks.} Schematic of SH polarizations from interfaces (blue arrows) and individual layers (red arrows) in \textbf{a},~$\langle 0^{\circ}, 30^{\circ}, 60^{\circ}\rangle$ and \textbf{b},~$\langle 0^{\circ}, 30^{\circ}, 0^{\circ}\rangle$ trilayer stacks. The interfacial SH polarizations interfere constructively and add up in \textbf{a,} but interfere destructively and cancel in \textbf{b}. Measured SHG signals under an incidence angle of $\beta=50^{\circ}$ ($\beta=-50^{\circ}$) are shown in red (blue) circles for \textbf{c},~$\langle 0^{\circ}, 30^{\circ}, 60^{\circ}\rangle$ and \textbf{d},~$\langle 0^{\circ}, 30^{\circ}, 0^{\circ}\rangle$ stacks. \textbf{e}, The difference of SHG intensity between incidence angles of $\beta = \pm 50^{\circ}$ shows vanishing (orange) and enhanced (green) interfacial susceptibility in the $\langle$0$^{\circ}, $30$^{\circ}, $0$^{\circ}\rangle$ and $\langle$0$^{\circ}, $30$^{\circ}, $60$^{\circ}\rangle$ stacks.
\label{Fig3}
\end{figure}

\begin{figure}[h]%
\centering
\includegraphics[width=1\textwidth]{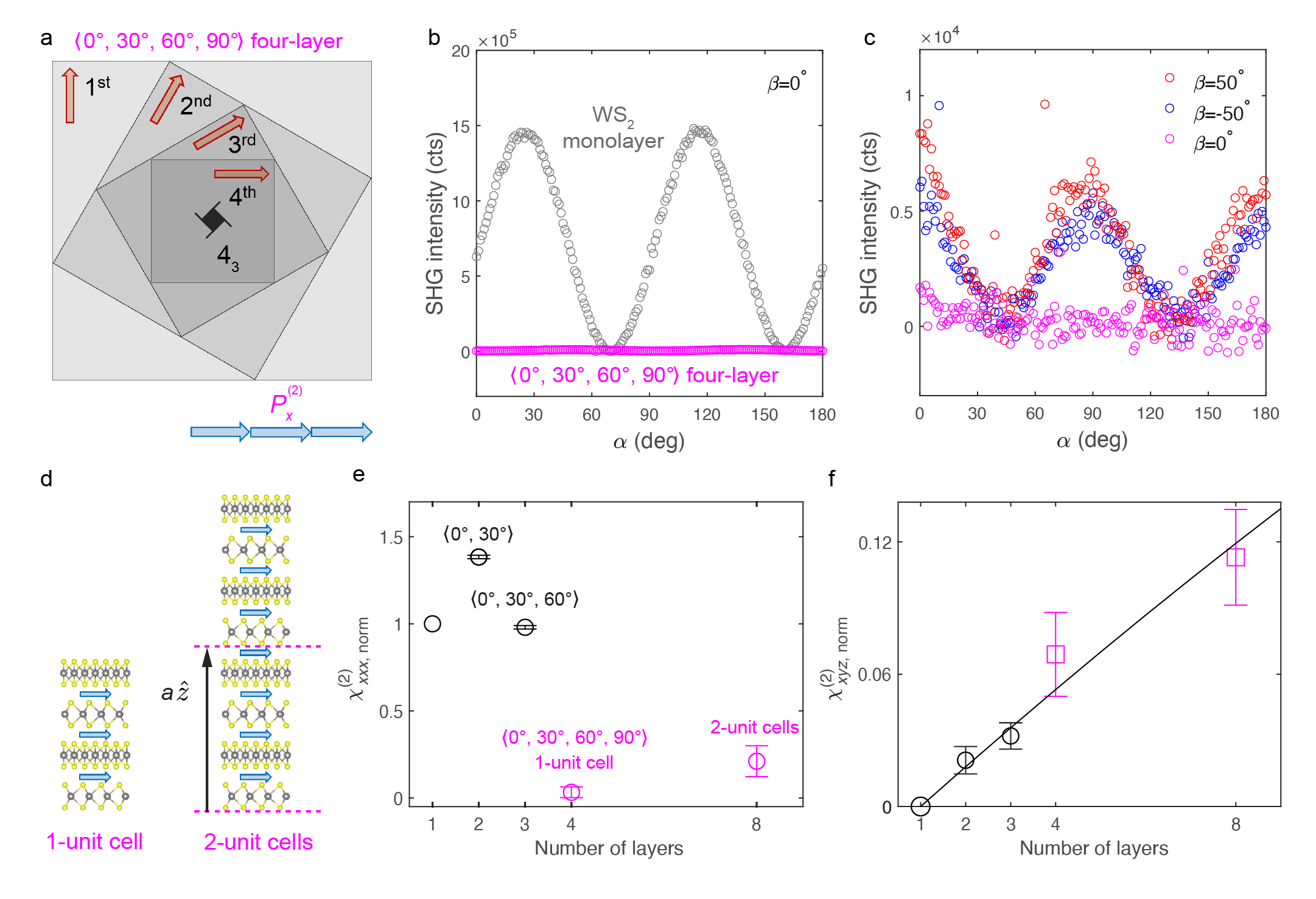}\\
\noindent\justifying\textbf{Fig.~4 $|$ Scalable nonlinear optical metamaterial. a,} A pseudo four-fold screw symmetry (4$_3$) in a four-layer WS$_2$ forbids in-plane SH polarizations (red arrows) while adding up all interfacial SH polarizations (blue arrows). \textbf{b}, SHG signals under normal excitation show significant suppression of in-plane responses in a four-layer stack (purple), compared to a monolayer (gray). \textbf{c}, SHG signals from the four-layer stack under the incidence angles of $-50^{\circ}$ (red), $0^{\circ}$ (purple), and $50^{\circ}$ (blue). \textbf{d}, Schematic of a twisted four-layer as a scalable metamaterial unit cell that can be stacked along the $\hat{\mathbf{z}}$ direction. Four- and eight-layers with screw symmetry have \textbf{e}, suppressed normalized in-plane susceptibility, $\chi^{(2)}_{xxx, {\rm norm}}$, and \textbf{f}, enhanced normalized interfacial susceptibility, $\chi^{(2)}_{xyz, {\rm norm}}$, linearly with the number of stack layers. The error bars represent standard deviations. See Methods for measurement details.
\label{Fig4}
\end{figure}

\begin{figure}[h]%
\centering
\includegraphics[width=0.8\textwidth]{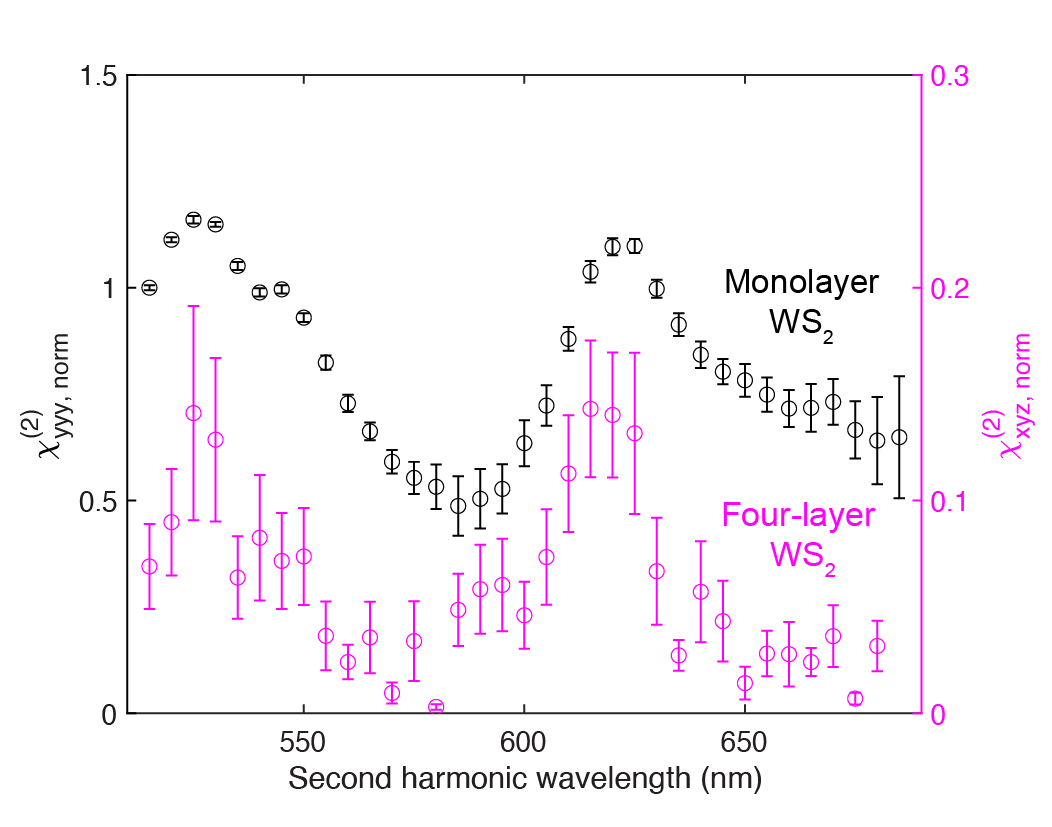}\\
\noindent\justifying\textbf{Fig.~5 $|$ Enhanced nonlinear susceptibility by exciton resonances.} Normalized in-plane susceptibility, $\chi^{(2)}_{yyy}$, from a monolayer (black circles) and interfacial susceptibility, $\chi^{(2)}_{yyy}$, from a twisted four-layer (purple circles), are both enhanced when the second-harmonic frequency is on resonance with excitons. The error bars represent the noise root mean square.
\label{Fig5}
\end{figure}

\clearpage

\section*{References}
\bibliographystyle{naturemag}
\bibliography{bibliography.bib}

\end{document}